\documentclass{elsart}

\usepackage[square,comma]{natbib}
\usepackage{graphicx}
\usepackage{pxfonts}
\usepackage{lineno}

\usepackage{amssymb}

\journal{}

\begin{document}

\thispagestyle{empty}
\begin{Large}
\textbf{DEUTSCHES ELEKTRONEN-SYNCHROTRON}

\textbf{\large{Ein Forschungszentrum der
Helmholtz-Gemeinschaft}\\}
\end{Large}

DESY 10-239

December 2010

\begin{eqnarray}
\nonumber &&\cr \nonumber && \cr \nonumber &&\cr
\end{eqnarray}
\begin{eqnarray}
\nonumber
\end{eqnarray}
\begin{center}
\begin{Large}
\textbf{Self-seeded operation of the LCLS hard X-ray FEL in the long-bunch mode}
\end{Large}
\begin{eqnarray}
\nonumber &&\cr \nonumber && \cr
\end{eqnarray}

\begin{large}
Gianluca Geloni,
\end{large}
\textsl{\\European XFEL GmbH, Hamburg}
\begin{large}

Vitali Kocharyan and Evgeni Saldin
\end{large}
\textsl{\\Deutsches Elektronen-Synchrotron DESY, Hamburg}
\begin{eqnarray}
\nonumber
\end{eqnarray}
\begin{eqnarray}
\nonumber
\end{eqnarray}
ISSN 0418-9833
\begin{eqnarray}
\nonumber
\end{eqnarray}
\begin{large}
\textbf{NOTKESTRASSE 85 - 22607 HAMBURG}
\end{large}
\end{center}
\clearpage
\newpage

\begin{frontmatter}



\title{Self-seeded operation of the LCLS hard X-ray FEL in the long-bunch mode}


\author[XFEL]{Gianluca Geloni\thanksref{corr},}
\thanks[corr]{Corresponding Author. E-mail address: gianluca.geloni@xfel.eu}
\author[DESY]{Vitali Kocharyan}
\author[DESY]{and Evgeni Saldin}

\address[XFEL]{European XFEL GmbH, Hamburg, Germany}
\address[DESY]{Deutsches Elektronen-Synchrotron (DESY), Hamburg,
Germany}

\begin{abstract}
Self-seeding options for the LCLS baseline were recently investigated using a scheme which relies on a single-crystal monochromator in Bragg-transmission geometry. The LCLS low-charge ($0.02$ nC) mode of operation was considered in order to demonstrate the feasibility of the proposed scheme. The wakefield effects from the linac and from the undulator vacuum chamber are much reduced at such low charge, and can be ignored. In this paper we extend our previous investigations to the case of the LCLS mode of operation with nominal charge. Based on the LCLS start-to-end simulation for an electron beam charge of $0.25$ nC,  and accounting for the wakefields from the undulator vacuum chamber we demonstrate that the same simplest self-seeding system (two undulators with a single-crystal monochromator in between) is appropriate not only for short (few femtosecond) bunches, but for longer bunches too.
\end{abstract}

%
%
%
\end{frontmatter}



\section{\label{sec:intro} Introduction}

The LCLS, the world's first hard X-ray FEL, has demonstrated SASE lasing and saturation at 0.15 nm \cite{LCLS2}. This success motivated the planning of a significant upgrade over the next several years \cite{FRIS}.

In \cite{OURY4}-\cite{OURY6}, self-seeding options for the LCLS baseline were investigated using a scheme relying on a single-crystal monochromator in Bragg-transmission geometry.  The Bragg crystal reflects a narrow band of X-rays, resulting in a ringing within the passband in the forward direction, which can be used to seed the second undulator. The chicane creates a transverse offset for the electrons, washes out previous electron-beam microbunching, and provides a tunable delay of the electron-bunch with respect to the radiation, so that the electron beam only interacts with the ringing tail of the X-ray pulse.

The LCLS enables both a low charge, short bunch mode of operation characterized by an electron beam charge of $0.02$ nC and a $1~\mu$m rms-long lasing part of the bunch, and a nominal charge, long bunch mode of operation, characterized by an electron beam charge of $0.25$ nC and a $10~\mu$m rms-long lasing part of the bunch. The wakefield effects from the linac and from the undulator vacuum chamber can be ignored in the case of the low charge mode of operation. Therefore for simplicity, in \cite{OURY4}-\cite{OURY6} we limited our consideration to the low charge mode of operation, in order to demonstrate the feasibility of our scheme.

\begin{figure}
\includegraphics[width=1.0\textwidth]{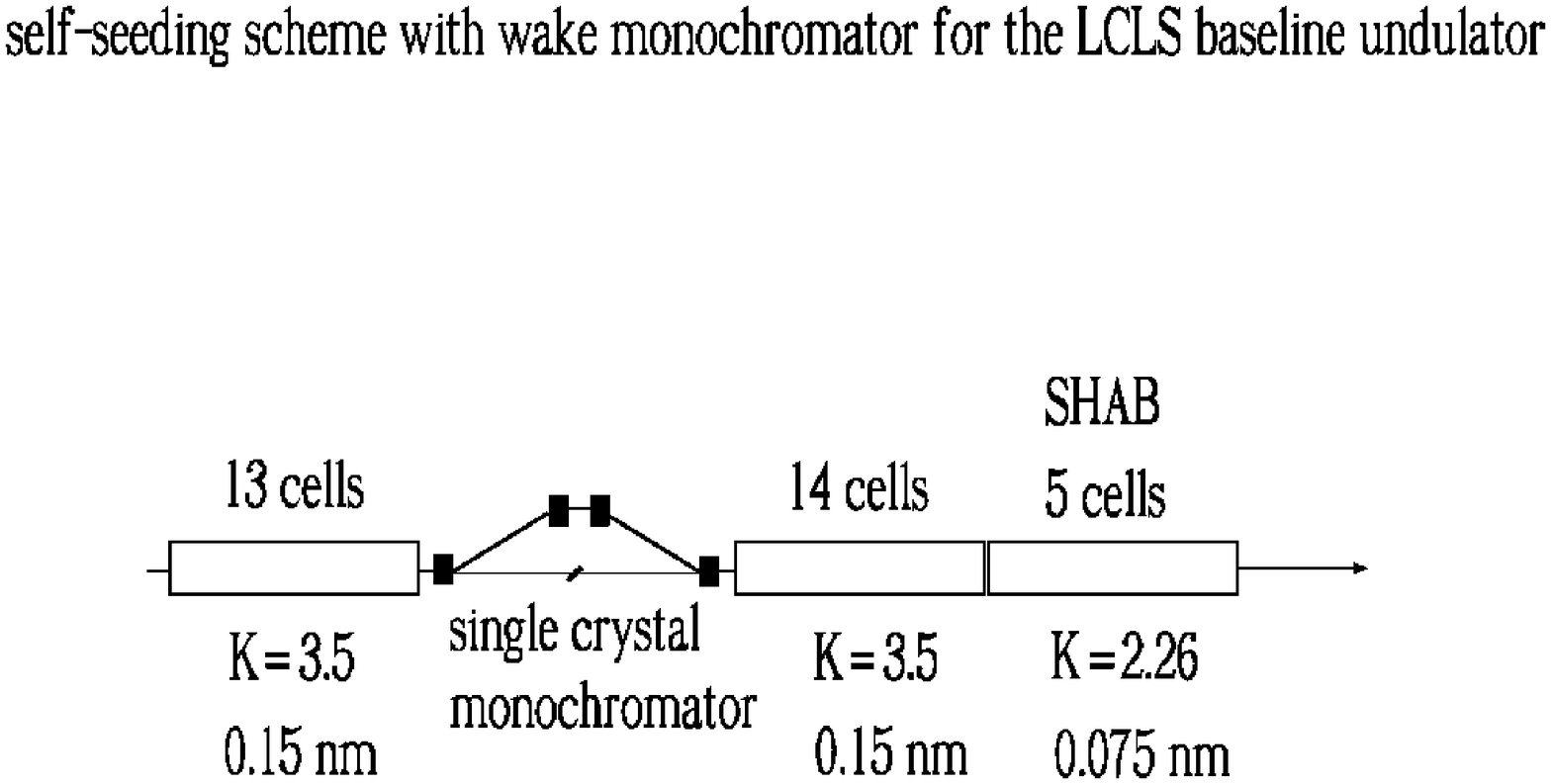}
\caption{Design of the LCLS baseline undulator system for generating highly
monochromatic hard X-ray pulses in the nominal charge (250 pC) mode of operation.} \label{lclslb1}
\end{figure}

\begin{figure}
\includegraphics[width=1.0\textwidth]{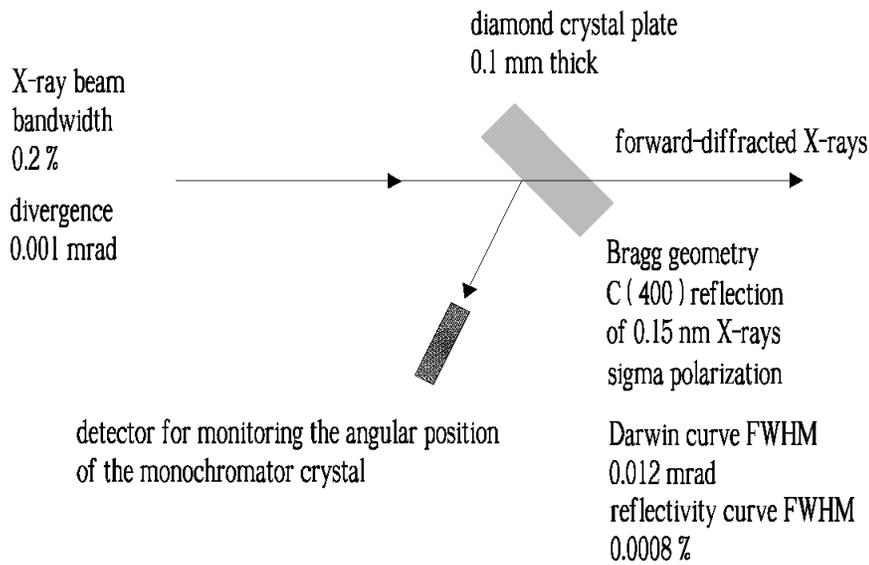}
\caption{Wake monochromator based on a $0.1$ mm -thick diamond crystal at 8 keV
for self-seeded operation of the LCLS baseline hard X-ray FEL. The Diamond (400) reflection is used.
} \label{lclslb2}
\end{figure}

\begin{figure}
\includegraphics[width=1.0\textwidth]{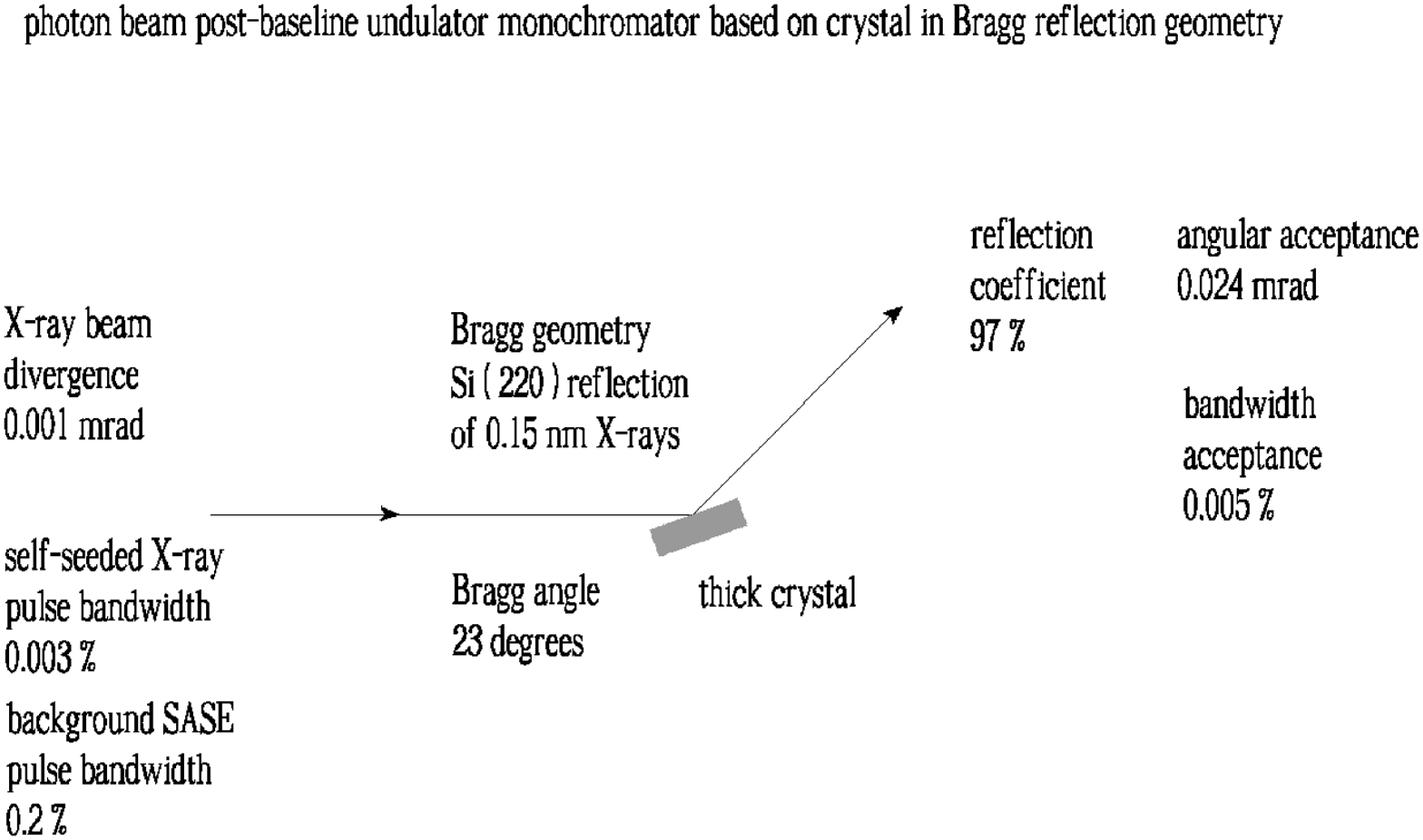}
\caption{Concept of post-baseline undulator monochromator for SASE background
suppression based on the use of a crystal in Bragg reflection geometry.} \label{lclslb3}
\end{figure}

In the present paper we extend our considerations to the nominal charge mode of operation for the LCLS. In particular, we propose a study of the performance of our self-seeding scheme for the LCLS, based on start-to-end simulations and accounting for the wakefields from the vacuum chamber. Particle tracking through the LCLS linac for the nominal charge case is presented in \cite{EMMA}. The resistive-wall wakefield in the LCLS undulator has been calculated including the frequency dependence of beam-pipe conductivity in
\cite{BANE}.  The wakefield generates an energy change, different for each time-slice along the bunch, shifting the slice out of resonance, and reducing the total FEL power. The expected performance  of the LCLS X-ray SASE FEL in the presence of these wakefields was studied in \cite{REIC,FAW}.   A  feasibility study of the two-bunch self-seeding scheme for LCLS nominal charge mode of operation is presented in \cite{DING}. We will consider results in \cite{DING} as the reference point for our investigations.

The setup which will be considered in this work is sketched in Fig. \ref{lclslb1}, and is composed of two undulator parts separated by a weak chicane. As in previous treatments \cite{OURY4}-\cite{OURY5}, the magnetic chicane accomplishes three tasks by itself.  It creates an offset for the single-crystal installation, it removes the electron micro-bunching produced in the first undulator, and it acts as a delay line between electrons and photons. The weak chicane should provide a delay in the order of $50 \mu$m,  meaning that $R_{56} \simeq 100 \mu$m or, to be conservative $R_{56} \simeq 200 \mu$m as in \cite{DING}, and can be installed within the space of a single undualtor segment ($4$ m long at the LCLS). Using a single crystal installed in the offset crated by the chicane, as in Fig. \ref{lclslb2}, it is possible to decrease the bandwidth of the radiation well beyond the XFEL design down to $3\cdot 10^{-5}$. Compared with the low charge case, where almost all the electrons radiate coherently, here we have a strong SASE signal due to the double-horn current profile and to the electron energy modulation generated by the undulator wakefields.
Nevertheless, as proposed in \cite{DING}, we can use a post-baseline monochromator for SASE background suppression, based on the use of a thick crystal in Bragg reflection geometry,  Fig. \ref{lclslb3}.

Following this introduction we will present a feasibility study for the LCLS, reviewing results in \cite{DING} and analysing the output from the setup in Fig. \ref{lclslb1}. We will show that the implementation of our scheme can yield, for the long bunch mode of operation, an output power in the order of $10-15$ GW and a bandwidth of $3\cdot 10^{-5}$.

\section{\label{sec:oper} Feasibility study }

In this Section we report on a feasibility study performed with the help of the FEL code GENESIS 1.3 \cite{GENE} running on a parallel machine. We will present a feasibility study for the long-pulse mode of operation of the LCLS, based on a statistical analysis consisting of $70$ runs. The overall beam parameters used in the simulations are as in \cite{DING}, and are presented in Table \ref{tt1}.

\begin{table}
\caption{Parameters for the low-charge mode of operation at LCLS used in
this paper.}

\begin{small}\begin{tabular}{ l c c}
\hline & ~ Units &  ~ \\ \hline
Undulator period      & mm                  & 30     \\
K parameter (rms)     & -                   & 2.466  \\
Wavelength            & nm                  & 1.55  \\
Energy                & GeV                 & 13.4 \\
Charge                & nC                  & 0.25\\
Bunch length (fw)     & $\mu$m              & 26.6  \\
Normalized emittance  & mm~mrad             & 0.4    \\
Energy spread         & MeV                 & 1.4   \\
\hline
\end{tabular}\end{small}
\label{tt1}
\end{table}

The long bunch mode of operation differs compared to the short one, in that the electron beam distribution, in terms of energy and current, is heavily influenced by wakes during the acceleration process, and in the undulator vacuum chamber.

The beam parameters at the entrance of the undulator are shown in Fig. 4 and Fig. 5  of \cite{DING}. The LCLS beam operated at nominal charge mode is characterized by a "double-horn" current distribution, which introduces additional energy modulation on the bunch due to wakefield effects in the undulator chamber. The  resistive-wall energy loss in the LCLS undulator  at nominal charge is shown in Fig. 6 of \cite{DING}. In Fig. \ref{current} we plot the electron beam current profile at the entrance of our setup, in Fig. \ref{lclslb1} we reproduce the electron beam energy profile again at the entrance of the setup, and in Fig. \ref{WakeU1} we reproduce the resistive wake in the undulator.

\begin{figure}[tb]
\includegraphics[width=1.0\textwidth]{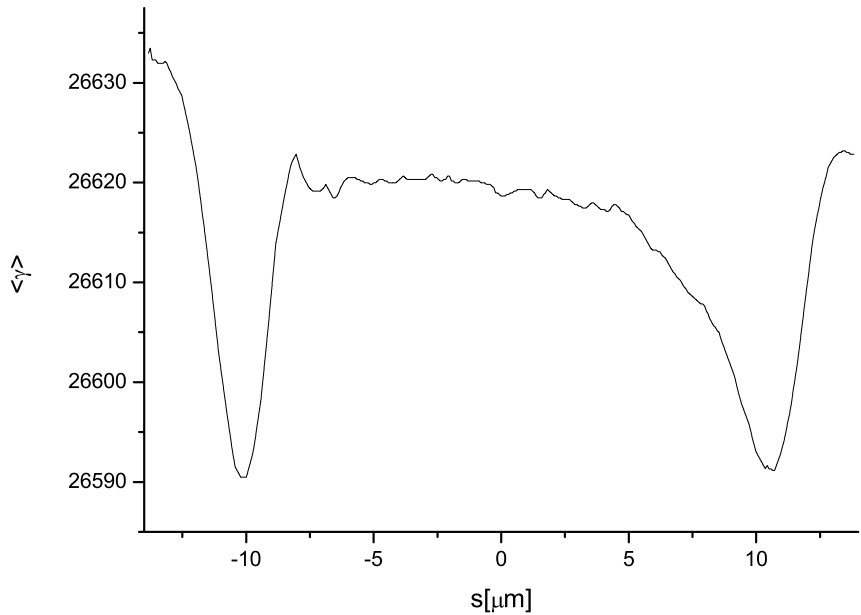}
\caption{Energy profile of the electron bunch at the entrance of the setup in Fig. \ref{lclslb1}, after \cite{DING}.} \label{EprofU1}
\end{figure}

\begin{figure}[tb]
\includegraphics[width=1.0\textwidth]{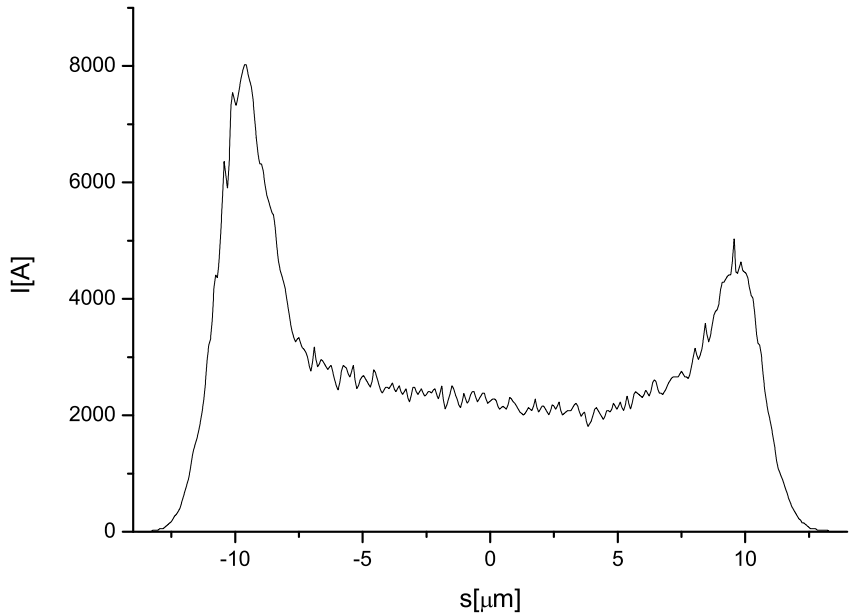}
\caption{Current profile of the electron bunch at the entrance of the setup in Fig. \ref{lclslb1}, after \cite{DING}} \label{current}
\end{figure}

\begin{figure}[tb]
\includegraphics[width=1.0\textwidth]{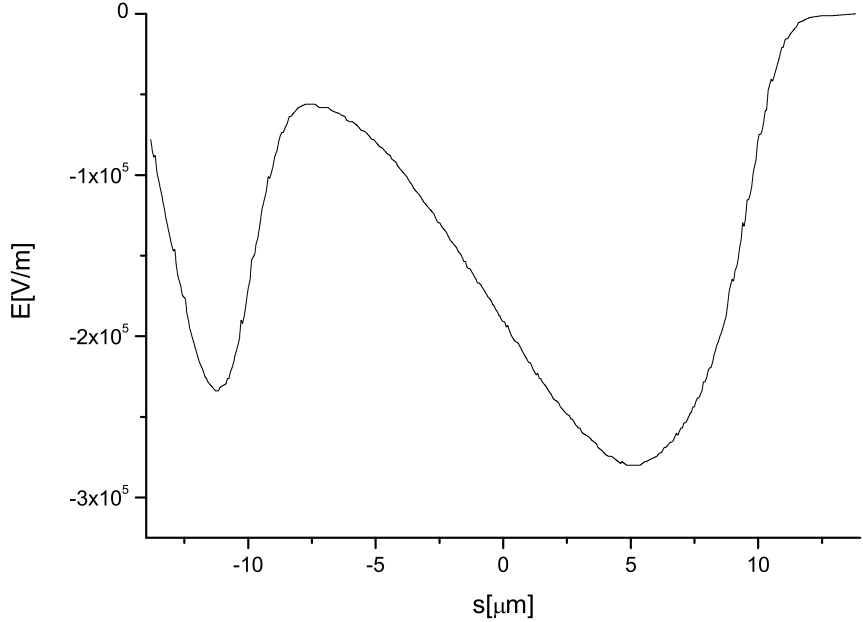}
\caption{Resistive wakefields in the LCLS undulator, after \cite{DING}. } \label{WakeU1}
\end{figure}

\subsection{First stage and crystal}

\begin{figure}[tb]
\includegraphics[width=1.0\textwidth]{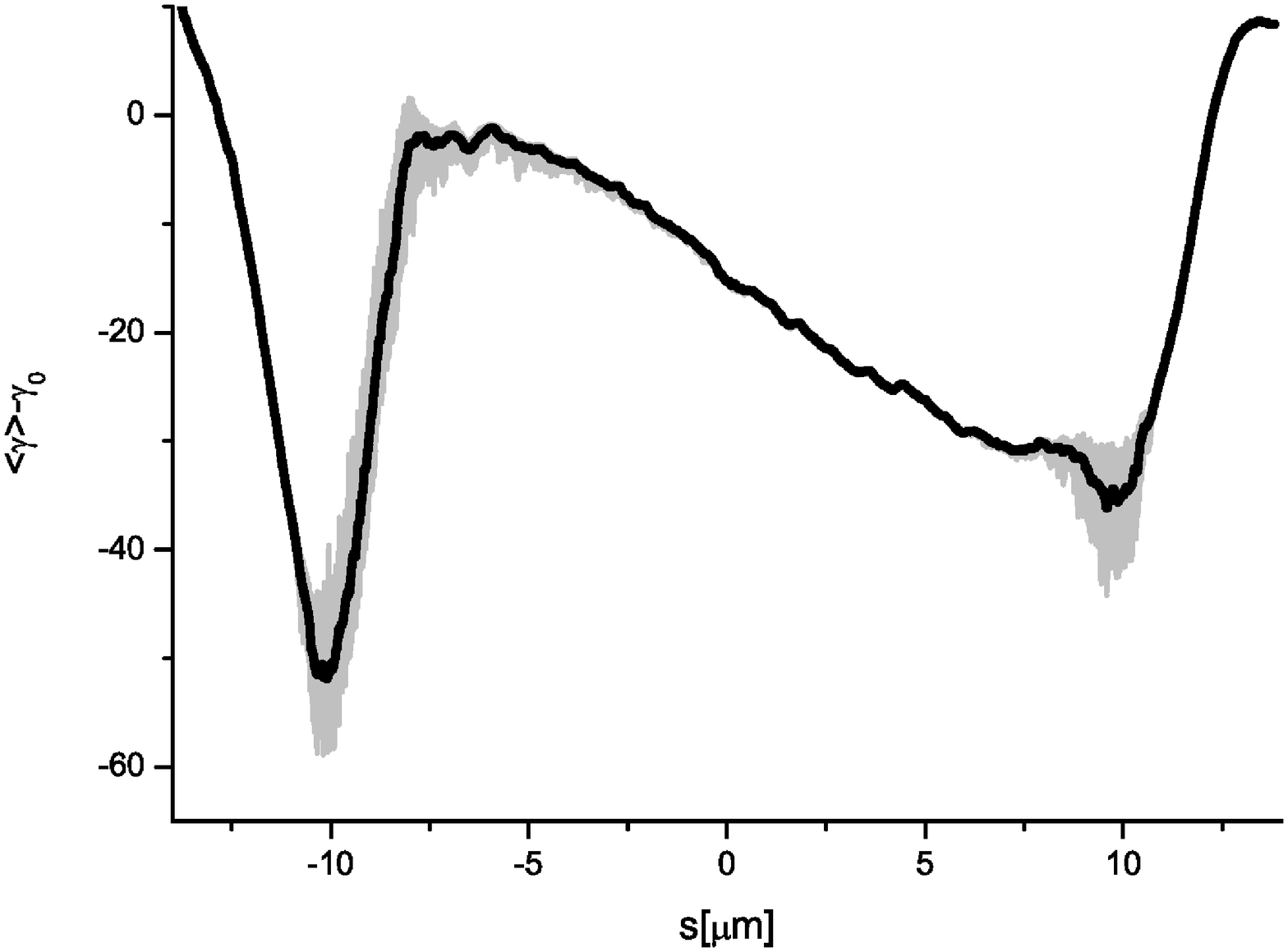}
\caption{Energy profile after the first SASE undulator (13 cells). Grey lines refer to single shot realizations, the black line refers to the average over 70 realizations.} \label{Penafter}
\end{figure}

\begin{figure}[tb]
\includegraphics[width=1.0\textwidth]{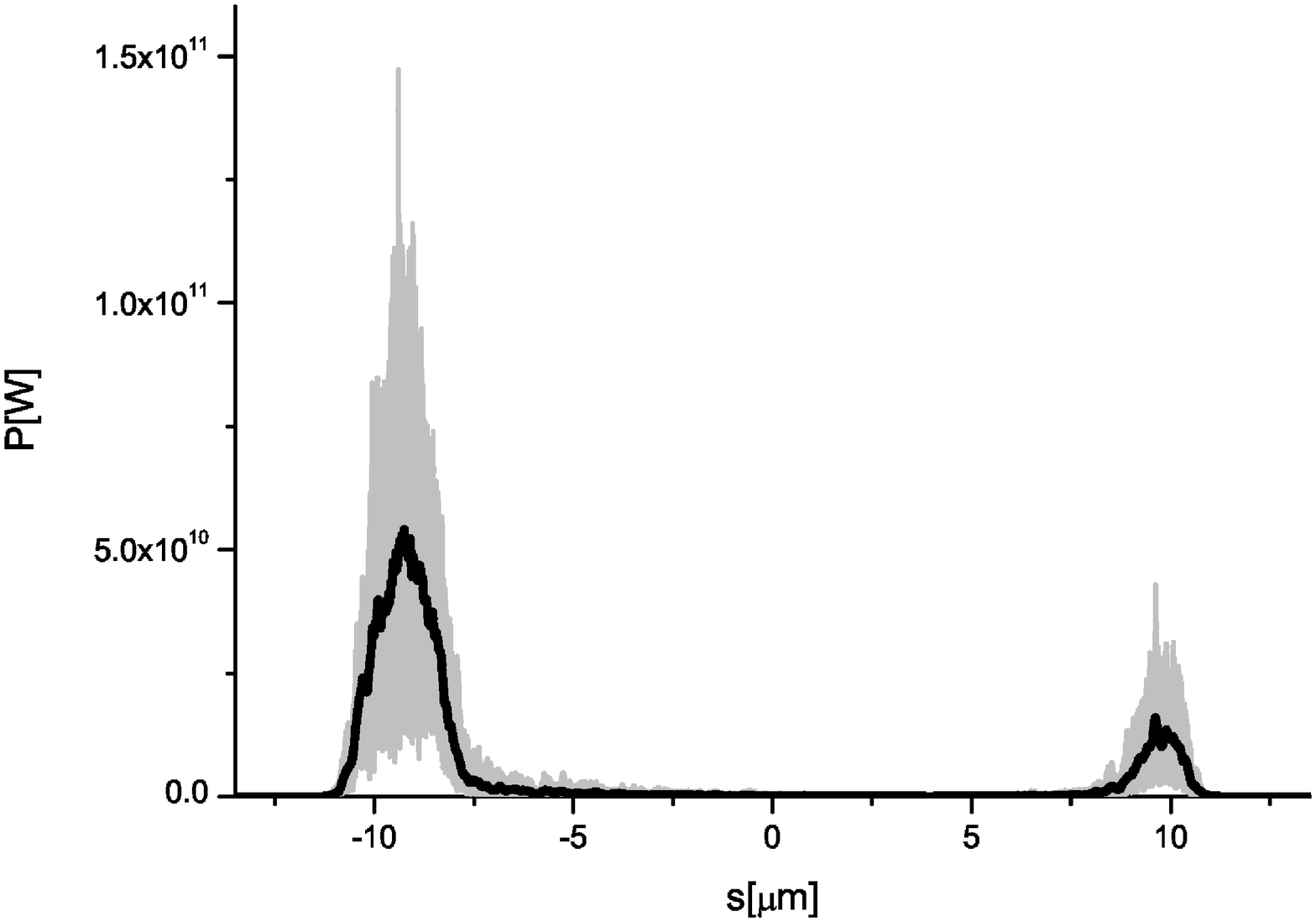}
\caption{Power distribution after the first SASE undulator (13 cells). Grey lines refer to single shot realizations, the black line refers to the average over 70 realizations.} \label{Pbefore}
\end{figure}

While the electron beam passes through the first undulator part shown in Fig. \ref{lclslb1}, it emits SASE radiation and it is subject to the resistive wake loss in the undulator vacuum chamber, see Fig. \ref{WakeU1}. In Fig. \ref{Penafter} we show different realizations of the electron beam energy profile (grey lines) and the average energy profile at the exit of the first undulator. Here we set the average betatron function $\beta=17.5$ m. At variance, in reference \cite{DING} a value $\beta=30$ m was used, which explains some of the differences in the output power after the first undulator compared the present article. The power before the crystal filter is shown in Fig. \ref{Pbefore}. Note the presence of two main radiation pulses, due to the "horns" in the current profile, Fig. \ref{current}, which are an effect of the wakes in the LCLS accelerator.

\begin{figure}[tb]
\includegraphics[width=1.0\textwidth]{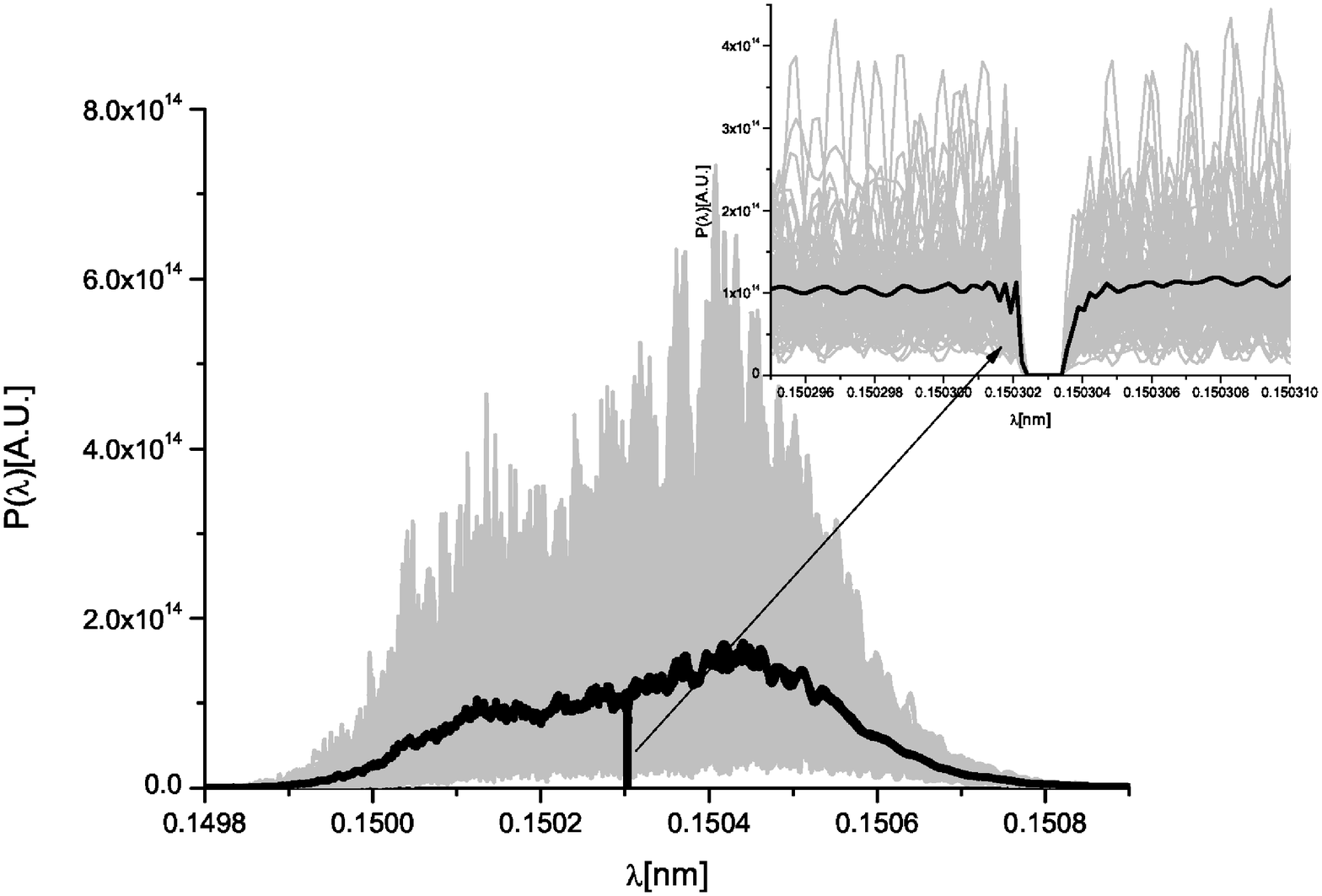}
\caption{Spectrum after the diamond crystals, $100~\mu$m-thick, C(400) reflection. The bandstop effect is clearly visible. Grey lines refer to single shot realizations, the black line refers to the average over 70 realizations.} \label{SPseed}
\end{figure}

As in previous works \cite{OURY4}-\cite{OURY6}, we use the C(400) reflection from a $100~\mu$m-thick diamond crystal in Bragg geometry, and we look at the transmitted beam. The band-stop effect of the filter is best shown in terms of spectrum, and it is evident by inspection of Fig. \ref{SPseed}, where such effect it is highlighted in the inset.

\begin{figure}[tb]
\includegraphics[width=1.0\textwidth]{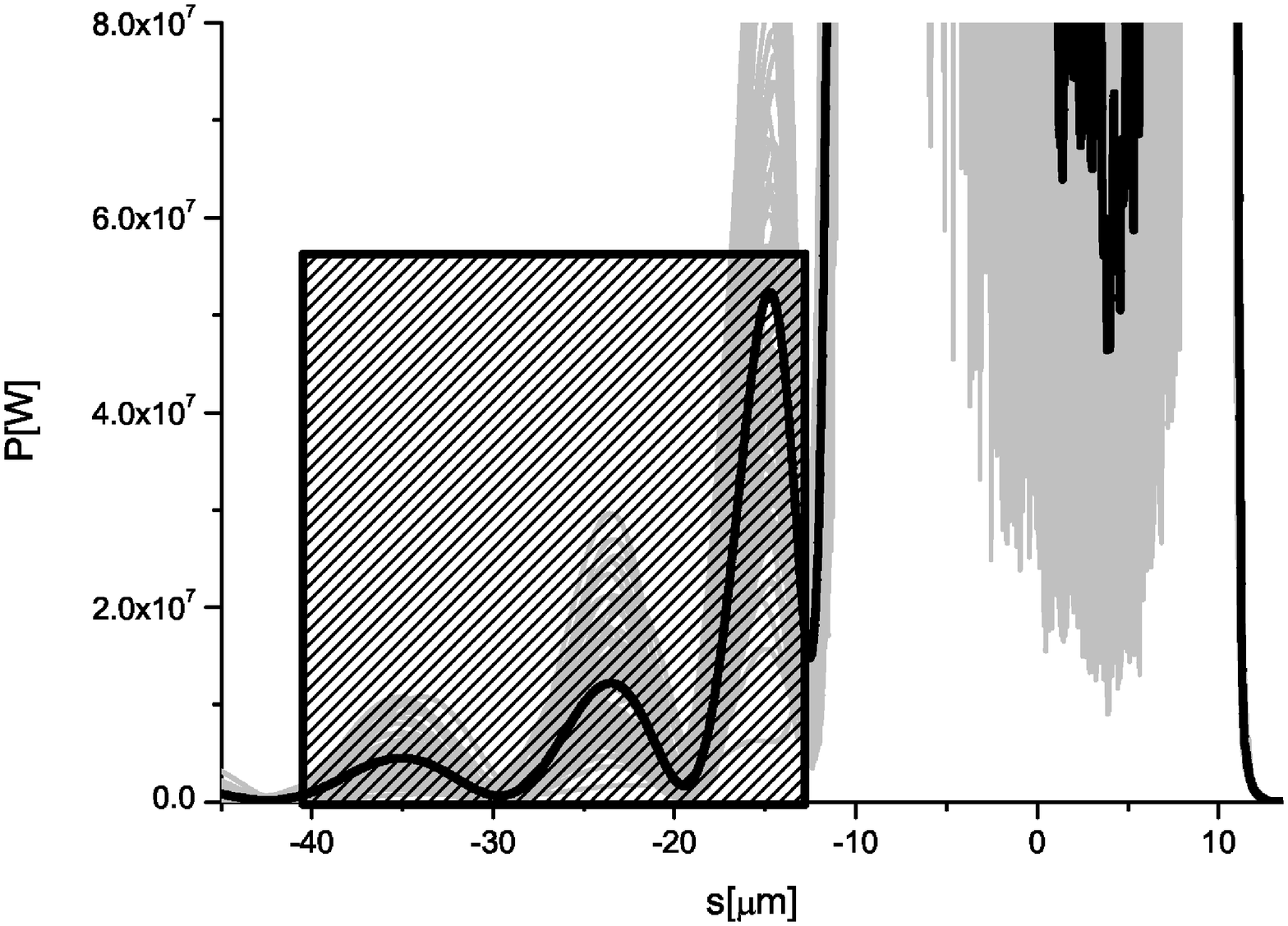}
\caption{Power distribution after the diamond crystals. The monochromatic tail due to the transmission through the bandstop filters is now evident on the left of the figure. Grey lines refer to single shot realizations, the black line refers to the average over 70 realizations. The square highlighted refers to the part of the monochromatic tail used for seeding the electron beam.} \label{seeded}
\end{figure}

In the time domain, the filtering operation through the Bragg crystal results in a ringing within the passband in the forward direction, which can be used to seed the electron beam in the second undulator. We show the ringing in Fig. \ref{seeded}, where we also highlight the portion of the radiation used for seeding the electron beam. Several oscillations are now used, because the bunch length is much longer than the inverse bandwidth of the crystal. It should be noted how the main contributions to the SASE radiation in Fig. \ref{Pbefore} are mainly due to the high-current "horns" at the beginning and at the end of the electron beam, Fig. \ref{current}. The relatively large seed power in the order of $5 - 20$ MW, is an effect of the strongly non Gaussian shape of the electron beam profile. Since the electron bunch is delayed of about $30 \mu$m\footnote{Actually we used a delay of $27~\mu$m. It should be noted that the stepped profile shape of the LCLS electron beam allows for such exact delay, while in the case of a Gaussian profile with the same FWHM, a much larger delay of about $60~\mu$m should be used.}, instead of about $6 \mu$m in the low charge mode of operation, one would expect a smaller seed power, compared to the low charge mode case. This apparent contradiction is explained by the presence of the large "horn" in the current profile, peaking at $8$ kA. In fact, the main horn in the LCLS electron beam reaches saturation after $13$ sections and is no more suitable for lasing. On the contrary, the flat part of the beam is still in the deep non-linear regime, and its quality is not wasted. In some sense, the strongly non Gaussian current configuration at the LCLS allows us to exploit a fresh-bunch technique where a part of the electron bunch is used to produce SASE (and, after filtering through the diamond crystal, the seeding signal), while another part is seeded and produces the final output.  The study of the implementation of a tapering schemes for the nominal charge mode of operation, as  done for the European XFEL \cite{OURY3} will be the subject of future work.

\subsection{Output characteristics}

\begin{figure}[tb]
\includegraphics[width=1.0\textwidth]{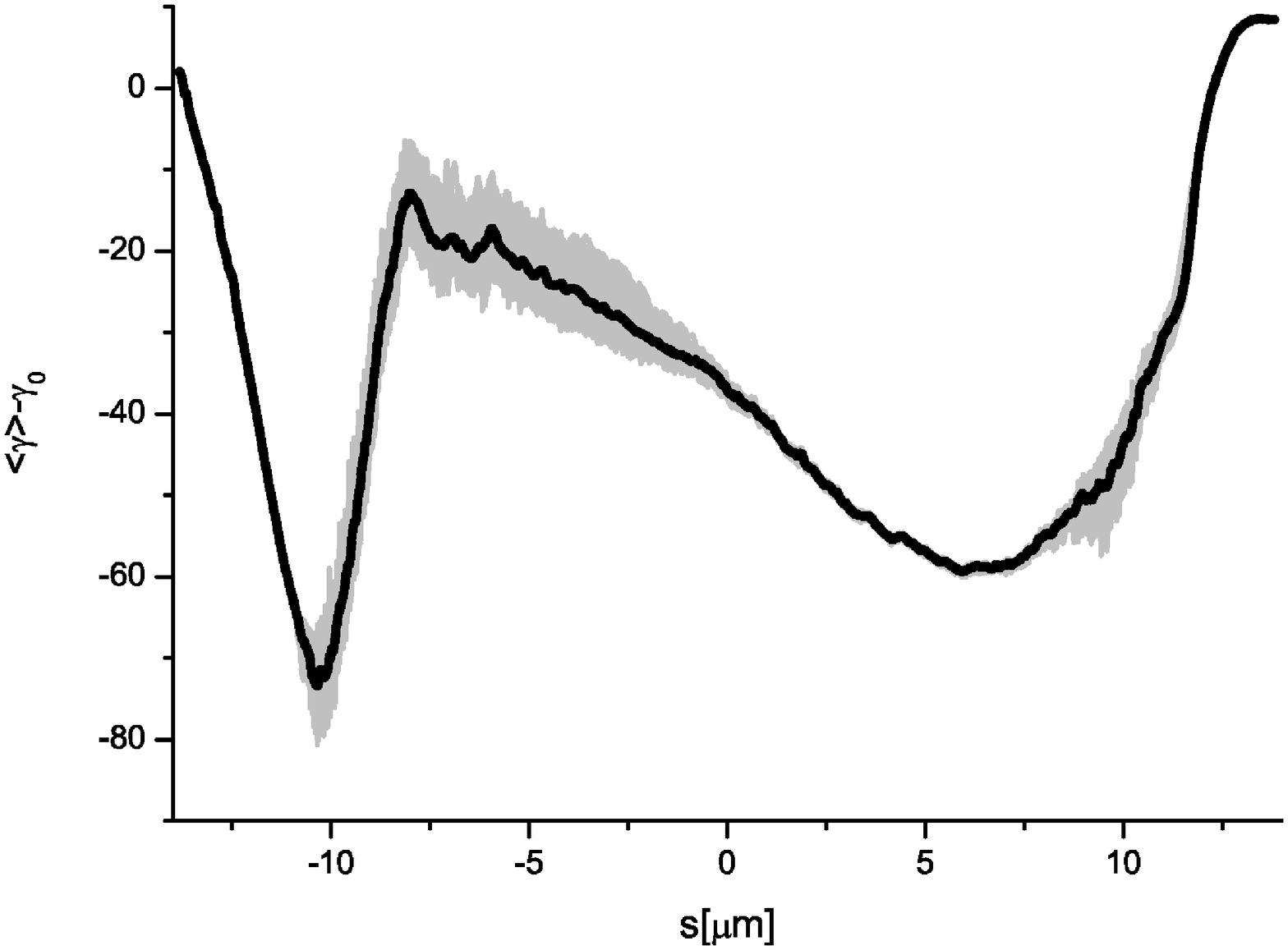}
\caption{Energy profile after the second SASE undulator (14 cells). Grey lines refer to single shot realizations, the black line refers to the average over 70 realizations.} \label{Penafter2}
\end{figure}

\begin{figure}[tb]
\includegraphics[width=1.0\textwidth]{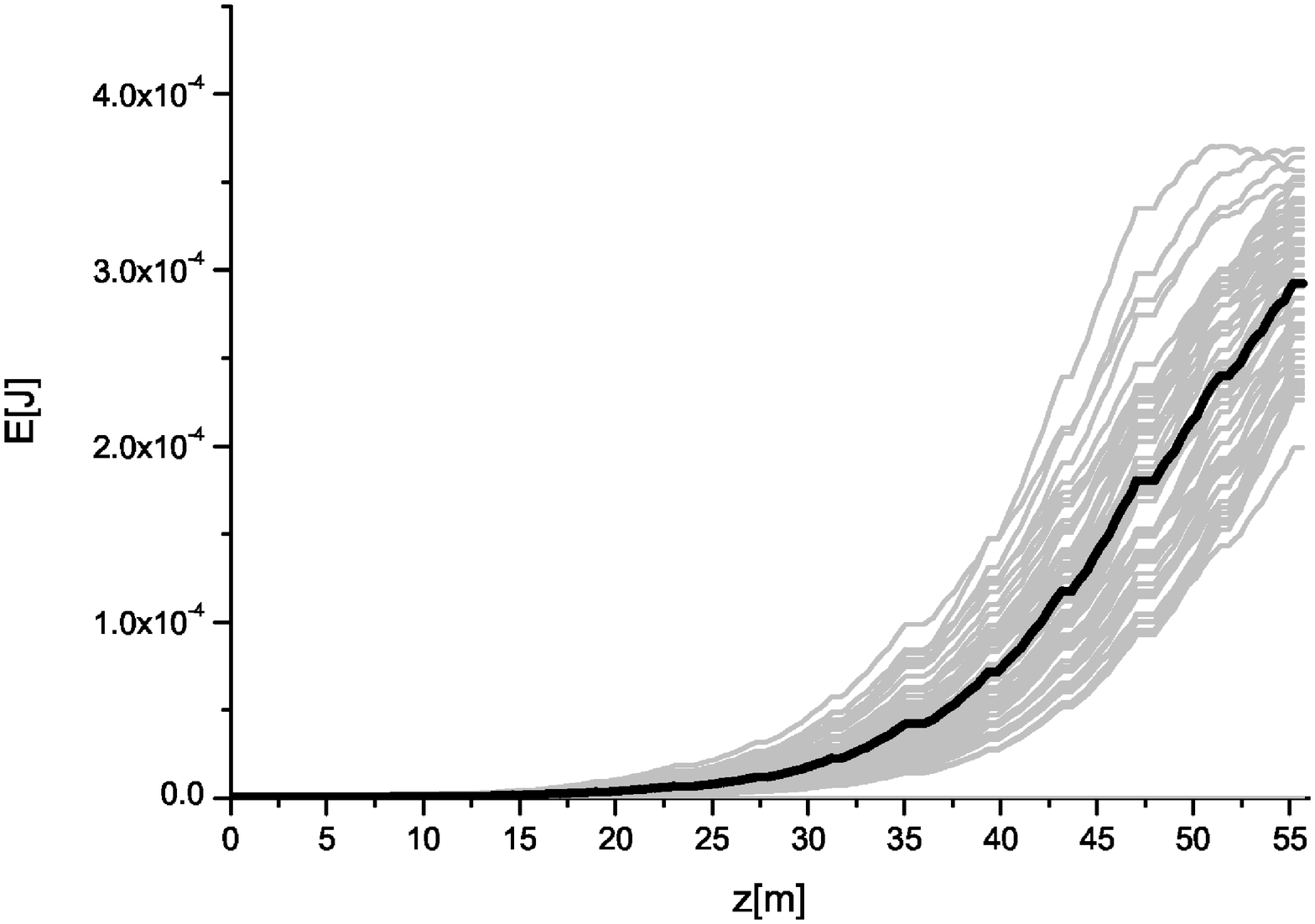}
\caption{Energy in the X-ray radiation pulse versus the length of the output undulator. Grey lines refer to single shot realizations, the black line refers to the average over fifty realizations.} \label{Enpout}
\end{figure}

\begin{figure}[tb]
\includegraphics[width=1.0\textwidth]{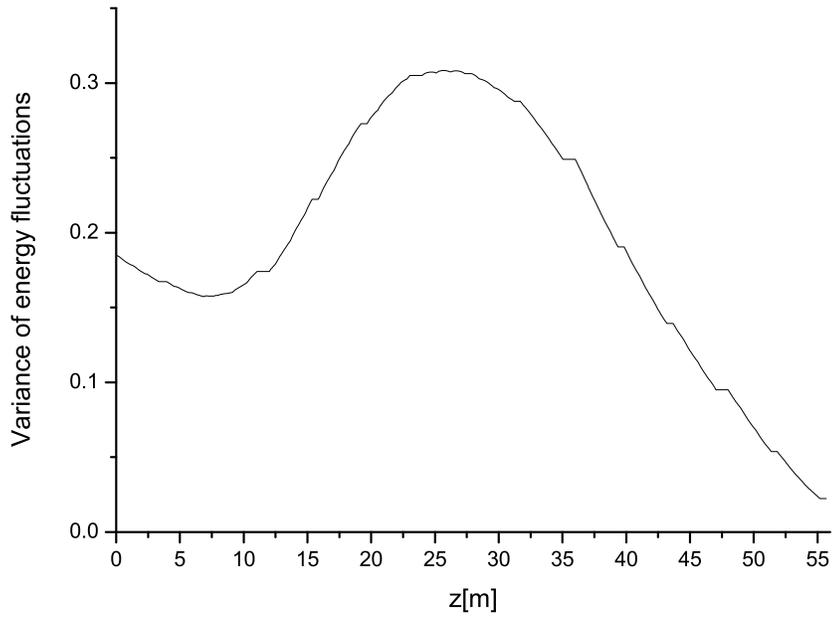}
\caption{Energy deviation from the average (variance) as a function of the distance inside the output undulator.} \label{variance}
\end{figure}

\begin{figure}[tb]
\includegraphics[width=1.0\textwidth]{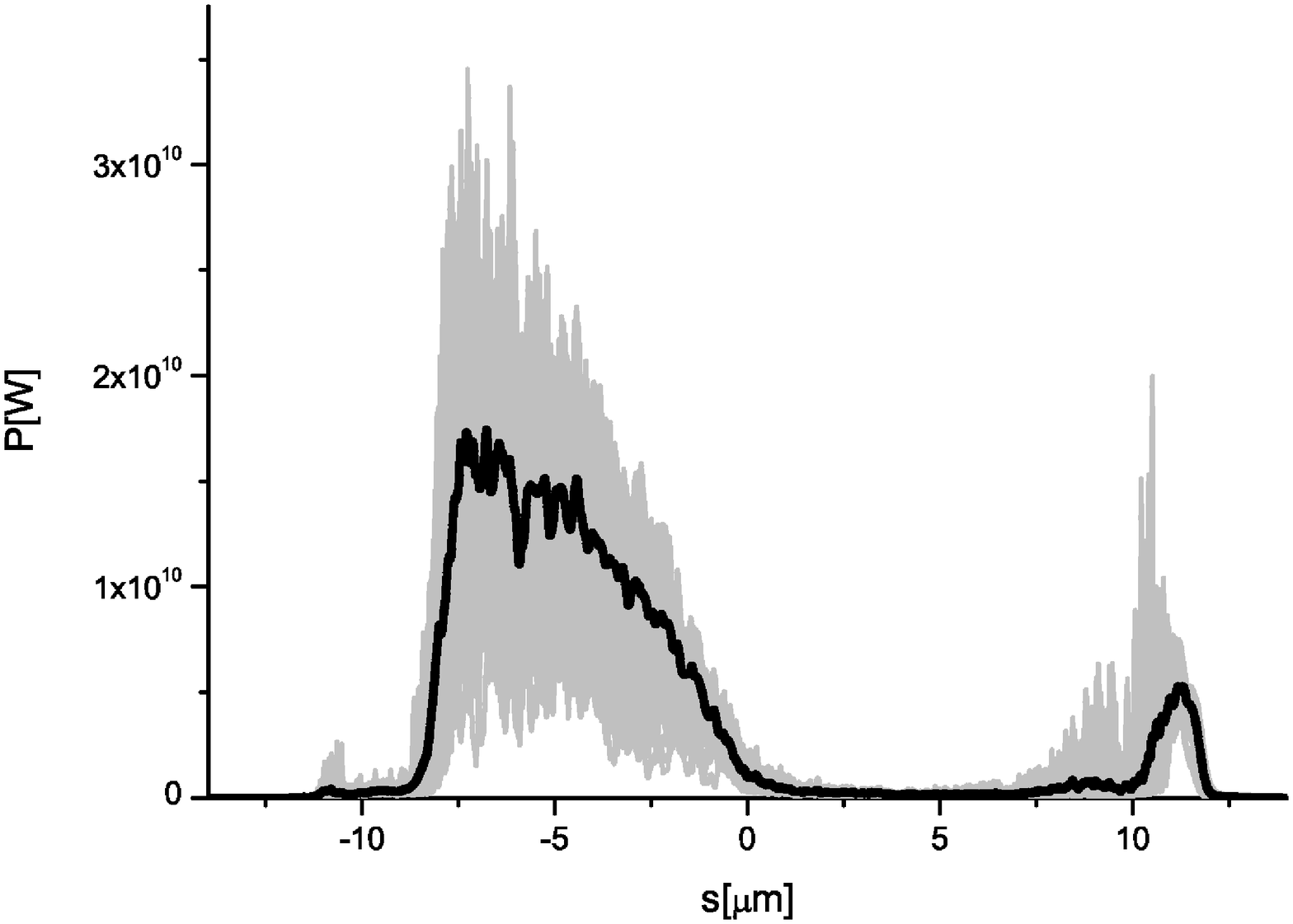}
\caption{Output power of the setup. Grey lines refer to single shot realizations, the black line refers to the average over 70 realizations.} \label{Pout}
\end{figure}

\begin{figure}[tb]
\includegraphics[width=1.0\textwidth]{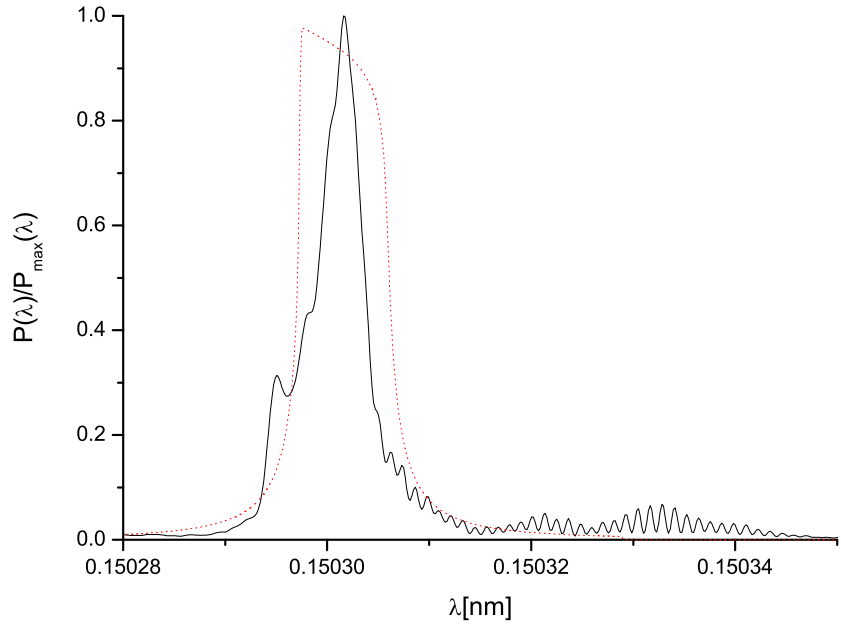}
\caption{(Solid line) Average output spectrum of the setup, normalized. (Dotted line) Shape of the thick crystal filter Si(220).} \label{Spout}
\end{figure}

\begin{figure}[tb]
\includegraphics[width=1.0\textwidth]{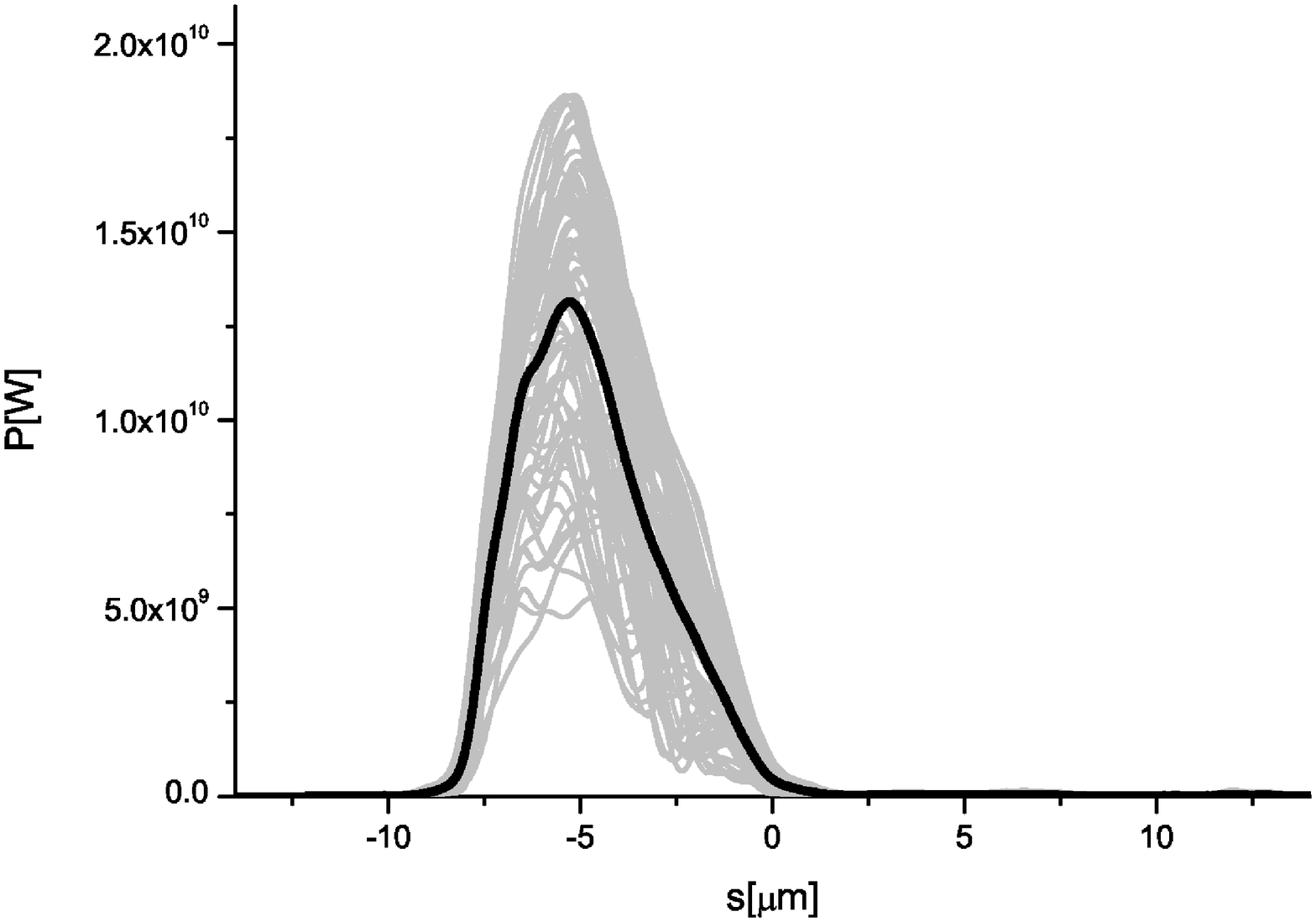}
\caption{Output power of the setup after filtering through a Si(22o) crystal. Grey lines refer to single shot realizations, the black line refers to the average over 70 realizations.} \label{Poutfilt}
\end{figure}

After the crystal, the electron beam and the seed are superimposed in the second undulator part. The energy profile of the electron bunch after the second undulator is shown in Fig. \ref{Penafter2}, where the effects of the resistive wake in the second undulator can be seen by comparison with Fig. \ref{Penafter}. As concern the output radiation characteristics, the energy per pulse as a function of the undulator length is shown in Fig. \ref{Enpout}. The energy variation, also as a function of the output undulator length, is presented in Fig. \ref{variance}. The output power of our setup is shown in Fig. \ref{Pout}. The corresponding average spectrum can be seen in Fig. \ref{Spout}. As one can see from the figures, our single-bunch self-seeding technique leads to the production of x-ray pulses with a relative bandwidth of about $3\cdot 10^{-5}$, yielding about $12$ GW power, and with $15$ fs FWHM duration, very similar to the output results from \cite{DING}, which is based on the two-bunch technique. The output signal can be further filtered to obtain better spectral properties. For example, in Fig. \ref{Spout} we show, together with the average spectrum, the spectral line of a thick Si(220) filter, which is characterized by a relative bandwidth of $5\cdot 10^{-5}$ FWHM, according to the post-baseline monochromatization scheme in Fig. \ref{lclslb2}. Such filtering procedure corresponds to the output power in Fig. \ref{Poutfilt}, to be compared with Fig. \ref{Pout}.


\section{\label{conc} Conclusions}

In this article we presented a study for our new monochromator scheme based on the use of a single crystal in Bragg geometry for the LCLS long bunch mode of operation, and demonstrated its feasibility. We accounted for wakefields in the accelerator and resistive wakes in the undulator chamber. The presence of these wakes yield a very peculiar current distribution, characterized by two very high-current "horns" at the bunch extremities, and a flat-top in between. These "horns" are shown to lase and saturate in the first part of the undulator, and are responsible for a large seeding signal, which keeps the background SASE noise to a small level. Such advantage, however, is compensated by the fact that tapering in the long bunch mode of operation becomes more complicated to implement. We demonstrated that the setup can yield, after a final filtering, an average pulse with $12$ GW peak power, a duration of $15$ fs, and a bandwith of $3\cdot 10^{-5}$. These characteristics, obtained using a single-bunch self-seeding scheme, are very similar to those reported in \cite{DING}, which makes use of the double bunch self-seeding scheme, where $9$ GW peak power and $10$ fs duration are reported. This is not surprising. Independently of the kind of monochromator used, in both cases we are taking advantage of a two-undulator scheme and we are working in the same conditions concerning undulator and electron beam properties. Since both monochromators provide sufficiently large seed compared with the shot noise power we same output characteristics are to be expected.

\section{Acknowledgements}

We are grateful to Massimo Altarelli, Reinhard Brinkmann, Serguei
Molodtsov and Edgar Weckert for their support and their interest
during the compilation of this work.

\end{document}